\documentclass{Rinton-P9x6}

\usepackage{amsmath}

\newcommand{\rmd}{{\rm d}}
\newcommand{\rme}{{\rm e}}
\newcommand{\rmi}{{\rm i}}
\newcommand{\bra}{\langle}
\newcommand{\ket}{\rangle}
\newcommand{\mbfe}{\mathbf{e}}
\newcommand{\mbfj}{\mathbf{j}}

\newcommand{\mbfr}{\mathbf{r}}

\newcommand{\mbfF}{\mathbf{F}}

\newcommand{\mbfH}{\mathbf{H}}

\newcommand{\newblock}{}
\begin{document}

\title{Matter waves from localized quantum sources}

\author{Tobias Kramer}

\address{Physik-Department T30,
Technische Universit\"at M\"unchen\\
James-Franck-Stra{\ss}e,
85747 Garching, Germany\\E-mail: tkramer@ph.tum.de
}

\maketitle

\abstracts{
Matter waves originating from a localized region in space appear commonly in physics. Examples are photo-electrons, ballistic electrons in nanotechnology devices (scanning-tunneling microscopy, quantum Hall effect), or atoms released from a coherent source (atom laser). We introduce the energy-dependent Green function as a suitable tool to calculate the arising currents. For some systems experimental data is available and in excellent agreement with the presented results.}

\section{Introduction}

The propagation of matter waves in external fields shows a wealth of interesting quantum-mechanical phenomena. Here, we only discuss uniform, static gravitational, electric, and magnetic fields. The corresponding Hamiltonians are quadratic polynomials in the momenta and position operators. Therefore all classical equations of motion are readily available, as well as the time-evolution operator in position space\cite{Grosche1998a} 
\begin{equation}
K(\mbfr,t|\mbfr',t_0)=\langle \mbfr|\exp(-\rmi\mbfH (t-t_0)/\hbar)|\mbfr'\rangle.
\end{equation}
However, the knowledge of the quantum-mechanical time-evolution is not sufficient to describe experiments in which the energy $E$ of the travelling particle is fixed, rather than the travel-time $t-t_0$. The energy-dependent counterpart of the time-evolution operator is the energy-dependent Green function, as we will see below. Surprisingly, analytic results for three-dimensional energy-dependent Green functions are scarce compared to the available time-dependent kernels: Commonly applied methods like the Feynman-path integral cannot be used.

If the process is to be treated time-independent or stationary, we have a system with a reservoir (or quantum source) that feeds particles into the external fields at a constant rate. This concept might be surprising in a quantum-mechanical context, since the Schr\"odinger equation implies a conservation of the probability current. However we will see that in the quantum-source approach the currents are conserved outside the emitting source region. This is just one advantage compared to another possible approach:
On can also solve the Schr\"odinger equation for a bound particle in the presence of an external field that enables the particle to tunnel from the originally bound state\cite{Kleber1994a}. Since the overall time dependence of a decaying state is given by $\exp(-\rmi E t/\hbar)$, a negative imaginary component of $E$ will lead to an exponentially decreasing probability current. While this feature is not unexpected in dealing with a decay-process, the resulting wave-function and probability current cannot be properly normalized. In the following we consider stationary currents which implies that we can use a real-valued energy-parameter. In connection with a quantum source we can define a meaningful probability current. Historically, the quantum-source approach was advocated by Schwinger\cite{Schwinger1973a}.

\section{Quantum Sources and Currents}

Here we merely state some consequences of the introduction of an inhomogeneous source term $\sigma(\mbfr)$ to the Schr\"odinger equation\cite{Bracher2003a,Kramer2003c}:
\begin{equation}
\left[ E-\mbfH \right] \psi_{\rm sc}(\mbfr) = \sigma(\mbfr).
\end{equation}
The simplest example of such a source term is the $\delta$-distribution, which describes matter waves originating from a point-source. A point source allows us to obtain a solution in the form of a Green-function $G(\mbfr,\mbfr';E)$
\begin{equation}
\left[ E-\mbfH \right] G(\mbfr,\mbfr';E) = \delta(\mbfr-\mbfr').
\end{equation}
In order to define the Green-function uniquely, we have to specify boundary conditions. In the following we impose the boundary-condition that we describe the emission of outgoing waves. The energy-dependent Green function is linked to the time-evolution operator by a Laplace transform:
\begin{equation}\label{eq:GreenEnergy1}
G(\mbfr,\mbfr';E)=-\frac{\rmi}{\hbar}\lim_{\eta\rightarrow 0_+}\int_0^\infty \rmd t\;
\left\bra \mbfr | U(t,t_0) | \mbfr' \right\ket \rme^{i E t/\hbar-\eta t/\hbar}.
\end{equation}
We obtain the scattering wave function assigned to the quantum source $\sigma(\mbfr)$ by a convolution integral:
\begin{equation}
\label{eq:Multi1.4}
\psi_{\rm sc}(\mathbf r) = \int {\rm d}^3r'\, G(\mathbf r,\mathbf r';E) 
\sigma(\mathbf r').
\end{equation}
A quantity of interest is the associated current, its spatial distribution and energy dependence. The current distribution is actually experimentally accessible, and we will discuss some examples. We define the current density distribution $\mathbf j(\mathbf r)$
\begin{equation}
\label{eq:currdens}
\mathbf j(\mathbf r) = \frac{\hbar}{m} \Im[\psi_{\rm sc}(\mathbf r)^* 
\boldsymbol\nabla \psi_{\rm sc}(\mathbf r)]-\frac{e \mathbf{A}(\mbfr)}{m}{|\psi_{\rm sc}(\mbfr)|}^2,
\end{equation}
where $\mathbf A(\mathbf r)$ denotes the vector potential. Integration of $\mathbf j(\mathbf r)$ over a surface enclosing $\sigma(\mathbf r)$ yields the total current $J(E)$ emitted by the source:
\begin{equation}
\label{eq:CurrentTotal}
J(E) = - \frac2\hbar \Im\left[ \int {\rm d}^3r \int {\rm d}^3r' \sigma(\mathbf r )^* G(\mathbf r,\mathbf r';E) \sigma(\mathbf r') \right].
\end{equation}
Connected to the appearance of a source for particles we have to modify the equation of continuity. Instead of $\nabla\cdot\mbfj(\mbfr)=0$, valid for a stationary system in the absence of sources, we now find:
\begin{equation}
\label{eq:Multi1.6}
\boldsymbol\nabla\cdot\mathbf j(\mathbf r) = 
- \frac2\hbar \Im\left[ \sigma(\mathbf r)^* \psi_{\rm sc}(\mathbf r) \right].
\end{equation}
Note, that outside the source region the current is a conserved quantity. 

\section{The twin-slit experiment}

\subsection{Material slits}

As a first example we address the propagation of particles from a point source in a field-free environment. We want to calculate the current distribution for a particle which is emitted from the source located at $\mbfr_A$ and arrives on the detector screen at $\mbfr_B$. In between, a wall with two holes at $\mbfr_{H_1}, \mbfr_{H_2}$ is placed. To simplify our considerations, we will assume a point source of particles. Since the double-slit configuration contains two possible particle paths, one from $\mbfr_A\rightarrow \mbfr_{H_1}\rightarrow \mbfr_B$ and the other one from $\mbfr_A\rightarrow \mbfr_{H_2}\rightarrow \mbfr_B$, we have to form a superposition of the two transition amplitudes
\begin{eqnarray}\label{eq:PropagatorTwoSlit1}
K_{\text{slit}}(\mbfr_B,T|\mbfr_A,0)
&=&\int_0^T\rmd t_1\; 
K_{\text{free}}(\mbfr_B,T|\mbfr_{H_1},t_1)\;K_{\text{free}}(\mbfr_{H_1},t_1|\mbfr_{A},0)
\\\notag
&+&\int_0^T\rmd t_2\; K_{\text{free}}(\mbfr_B,T|\mbfr_{H_2},t_2)\;K_{\text{free}}(\mbfr_{H_2},t_2|\mbfr_{A},0).
\end{eqnarray}
By integrating over $t_1$ and $t_2$ from $0$ to $T$ we consider a time evolution that starts at $t=0$ at $\mbfr_A$ and ends at $t=T$ at $\mbfr_B$. In between a passage through the holes occurs at some intermediate time $t_i$. Knowing the form of the propagator $K_{\text{free}}$, we have to carry out the intermediate time integration over $t_i$. However, in an experimental setup we would like to eliminate the need to measure exactly at time $T$. Instead one prefers to record a static pattern on the detector screen for particles emitted with a fixed energy $E$. Switching to the energy-dependent Green function we obtain
\begin{multline}\label{eq:GTwoSlitInt}
G_{\text{slit}}(\mbfr_A,\mbfr_B;E)
=-\frac{\rmi}{\hbar}\int_0^\infty \rmd T\;\rme^{-\rmi E T/\hbar} K_{\text{slit}}(\mbfr_A,T|\mbfr_B,0)\\
=-\frac{\rmi}{\hbar}\int_0^\infty \rmd T\;\rme^{-\rmi E T/\hbar} \sum_{i=1}^2
\int_0^T\rmd t_i\; K_{\text{free}}(\mbfr_B,T|\mbfr_{H_i},t_i)\;K_{\text{free}}(\mbfr_{H_i},t_i|\mbfr_A,0)
\end{multline}
Using the convolution theorem for the Laplace transform\cite{Abramowitz1965a}, 
we can rewrite the Green function as
\begin{eqnarray}\label{eq:GTwoSlit}
G_{\text{slit}}(\mbfr_A,\mbfr_B;E)
&=&\rmi\hbar\sum_{i=1}^2 G_{\text{free}}(\mbfr_B,\mbfr_{H_i};E)\;G_{\text{free}}(\mbfr_{H_i},\mbfr_A;E).
\end{eqnarray}
The last form clearly shows the superposition of two scattering waves. The corresponding current density distribution shows the familiar interference pattern, which was observed in a twin-slit experiment with electrons\cite{Moellenstedt1959a}.

\subsection{A virtual twin-slit provided by a homogeneous force field}

In 1981 it was pointed out that a linear force field can act as a double slit without the need for an actual material slit\cite{Demkov1982a,Fabrikant1981a}. An extended analysis in the language of quantum propagators is available\cite{Bracher1999a,Kramer2002a}. The time-dependent propagator for a particle in a uniform force field $\mbfF=F\mbfe_z$ is given by
\begin{figure}[t]
\begin{center}
\epsfxsize=14pc 
\epsfbox{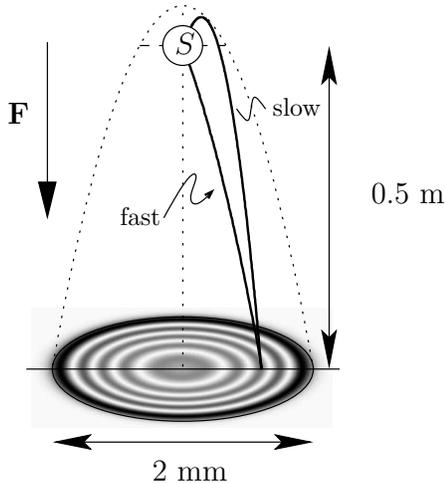}
\end{center}
\caption{
Trajectories in the uniform force field $\mbfF$. The dashed parabola denotes the classical accessible region. In this region, two trajectories with the same energy connect the source $S$ with a point on the detector plane. For electrons emitted in an electric field macroscopically visible interference fringes emerge.\label{fig:FieldSlit}}
\end{figure}
\begin{equation}\label{eq:Kfield}
K_{\text{field}}(\mbfr,T|\mbfr',0)
={\left(\frac{m}{2\pi\rmi\hbar T}\right)}^{3/2}
\exp\left(\frac{\rmi m {|\mbfr-\mbfr'|}^2}{2\hbar T}
         +\frac{\rmi F T}{2\hbar} (z+z')
	 -\frac{\rmi F^2 T^3}{24m\hbar}\right),
\end{equation}
where $T=t-t'$. As before, we want to calculate the probability amplitude for a particle traveling from the source $\mbfr_A$ to hit the detector at $\mbfr_B$. Using again a monochromatic particle source, the stationary probability amplitude is derived similarly to (\ref{eq:GTwoSlitInt}) and reads
\begin{equation}\label{eq:GfieldInt}
G_{\text{field}}(\mbfr,\mbfr';E)=-\frac{\rmi}{\hbar}\int_0^\infty \rmd T\;
\rme^{\rmi E T/\hbar}
K_{\text{field}}(\mbfr,T|\mbfr',0).
\end{equation}
Although $G_{\text{field}}(\mbfr,\mbfr';E)$ is available in closed analytic form\cite{Dalidchik1976a}, we will retain the integral form for the following discussion. We can approximate the integral with the method of stationary phases for a wide range of parameters. At a stationary point the derivative with respect to $T$ of the exponent in (\ref{eq:GfieldInt}) vanishes and we get a large contribution from the region close to this point to the otherwise oscillatory integral. The resulting biquadratic expression in $T$ has two roots for classically allowed motion $E>0$, and thus we obtain two real-valued solutions for the time of flight $T_{i}$. Galilei\cite{Galilei1638a} noticed this in his studies of the parabolic motion of free falling masses. The two classical trajectories produce the pronounced interference fringes depicted in Fig.~\ref{fig:FieldSlit}. The spacing of the fringes depends on the energy of the emitted particles $E$, the strength of the uniform force field $F$ and the distance between source and detector $(r,z)$. Following the idea of Demkov et~al.\cite{Demkov1982a,Fabrikant1981a}, Blondel constructed a device to realize the field double-slit (also for electrons), the so-called photodetachment-microscope\cite{Blondel1996a,Blondel1999a}. The photo-detachment is a high-precision spectrometer for the determination of the energy parameter and is used to determine the electron affinity of negative ions with extreme accuracy.

\section{Uncertainty principle in parallel magnetic and electric fields}

\begin{figure}[t]
\begin{center}
\epsfxsize=15pc 
\epsfbox{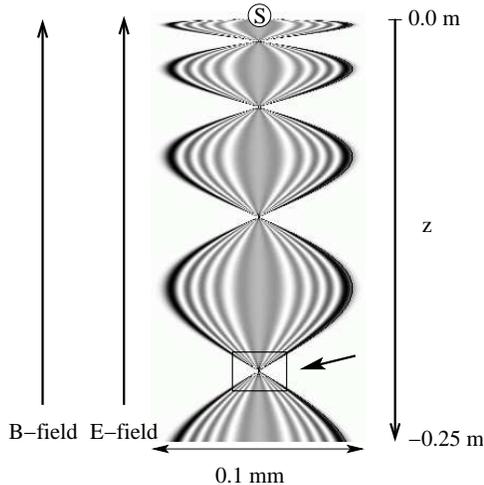}
\end{center}
\caption{
Plot of the current density distribution for parallel electric and magnetic fields. Parameters: Emission energy: $E=60.8$~$\mu$eV, electric field: $F=116$~eV/m, magnetic field: $B=0.001$~T. There is rotational symmetry about the $z$-axis.\label{fig:EBLense}}
\end{figure}

The addition of a homogeneous magnetic field to an electric field leads to additional classical trajectories that connect the source with a given point on the detector\cite{Kramer2001a}. A parallel magnetic field will enforces a cyclotron motion of the emitted electrons and establishes a lateral confinement (see Fig.~\ref{fig:EBLense}).
%
 
For low energies the classically allowed paths are confined closely to the path leading directly downwards from the quantum-source. In quantum mechanics, the focussing of the electrons due to the electromagnetic fields cannot be perfect, since Heisenberg's uncertainty principle enforces a minimum lateral distribution in position space along the escape path. Also negative energies $E<0$ lead to a tunneling current for which no classical trajectory exists. While the exact solution in terms of the energy-dependent Green function is still available, also a heuristic description in terms of the minimum uncertainty principle is possible\cite{Bracher1998a} and can be used to establish a tunneling time. 

\section{Coherent atomic ensembles and the atom laser}

The quantum-source approach is not limited to point-like emitters. Analytic solutions are available for a Gaussian source in an external homogeneous force field\cite{Kramer2002a}. A Bose-Einstein condensate (BEC) is a possible realization of a macroscopic quantum source.  The controlled and coherent release of atoms from such a condensate is called atom-laser.
\begin{figure}[t]
\begin{center}
\epsfxsize=27pc 
\epsfbox{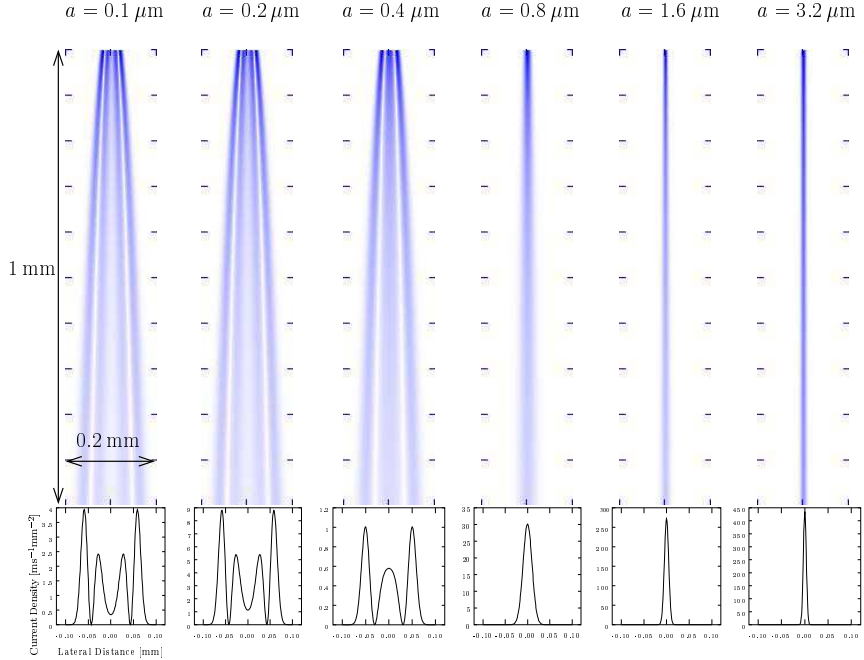}
\end{center}
\caption{
Transition from a strongly confined condensate to a more extended source distribution. There is rotational symmetry about the vertical axis. The source width is denoted by $a$. Interference fringes are clearly seen for $a\le0.4~\mu$m. Parameter: $\Delta\nu=E/(2\pi\hbar)=2.5$~kHz, $F=m_{\text{Rb}}\,g$, with $g=9.81$~m/s$^2$, and $m_{\text{Rb}}=87$~u.\label{fig:AtomLaser}}
\end{figure}
In\cite{Kramer2002a,Bracher2003a} we establish a completely analytical model for the description of the efficiency of such an atom laser and also for the density distribution in the resulting atom beam. Especially for strongly confined Bose-Einstein condensates (realized in so-called microtraps) the appearance of strong interference phenomena in the atomic beam is predicted. Remarkably, in the far-field sector in a uniform force field the Gaussian source may be replaced by a virtual point source which is shifted upwards in the external field. Thus we can readily apply the analysis of the virtual twin-slit in the previous section. However, the shift in position affects the energy-parameter of the Green-function and leads to effective negative energies. Therefore, for larger condensates the interference fringes are suppressed (see Fig.~\ref{fig:AtomLaser}).

\section{Electromagnetic fields and the density of states}

\begin{figure}[t]
\begin{center}
\epsfxsize=20pc 
\epsfbox{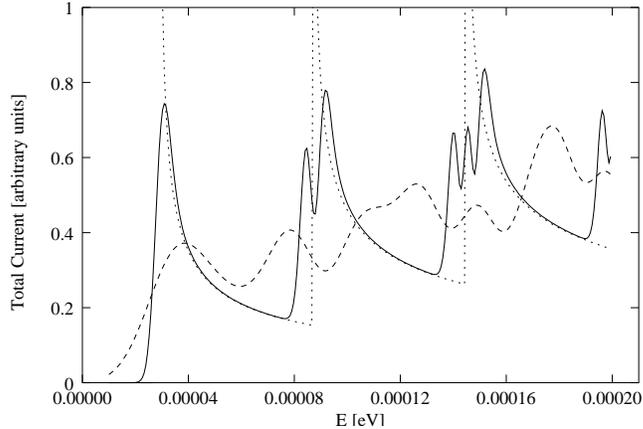}
\end{center}
\caption{
The transition from a weak electric field to a stronger electric field is shown. Parameter: Magnetic field; $B=0.5$~T, parallel electric field: $\mbfF_z=0$, perpendicular electric field: $\mbfF_y=(1,100,400)$~eV/m (dotted, solid, and dashed line respectively.
\label{fig:DOS}}
\end{figure}

Another use of quantum sources is the field of solid-state physics. There, considerable interest exists in expressions for the density of states (DOS) of a system. The DOS is defined by 
\begin{equation}
n(\mbfr;E)
=\langle\mbfr|\delta(E-\mbfH)|\mbfr\rangle
=-\frac{1}{\pi}\Im[G(\mbfr,\mbfr;E)].
\end{equation}
In our approach the DOS is directly proportional to the total current generated by a point-source, which is an surface-integral over the spatial current-distribution around the source. For the spatial current, classical trajectories can be compared with the quantum solution and show the difficult transition between a semi-classical and purely quantum-mechanical regime. In Fig.~\ref{fig:DOS} we demonstrate the influence of external fields on the DOS.

\section{Outlook and conclusions}

Quantum sources and the energy dependent Green functions provide an excellent tool for the analysis of stationary propagation in external fields. It is possible to extend the theory to include sources with an angular momentum (i.e.\ rotating BECs, or electrons coming from a specific atomic orbital)\cite{Bracher2003a}. In solid state physics, we can use the source-formalism to describe the electron propagation in a two-dimensional quantum-Hall system\cite{Kramer2003a}. These examples surely present only a glimpse of the possible applications of the source-framework\cite{Kramer2003b}.

\section*{Acknowledgments}

The author would like to thank H.~Moya-Cessa and V.I.~Man'ko for the invitation to present this work. C.~Bracher, M.~Kleber, and P.~Kramer contributed to this work by numerous helpful discussions. Financial support by the Leonhard-Lorenz foundation is gratefully acknowledged.


\begin{thebibliography}{10}

\bibitem{Grosche1998a}
C.~Grosche and F.~Steiner.
\newblock {\em Handbook of Feynman Path Integrals}, volume 145 of {\em Springer
  Tracts in Modern Physics}.
\newblock Springer, Berlin, 1998.

\bibitem{Kleber1994a}
M.~Kleber.
\newblock {\em Physics Reports}, 236(6):331, 1994.

\bibitem{Schwinger1973a}
J.~Schwinger.
\newblock {\em Particles, Sources, and Fields}, volume~2.
\newblock Addison-Wesley, 1973.

\bibitem{Bracher2003a}
C.~Bracher {\em et~al}.
\newblock {\em Phys.~Rev.~A}, 67:043601--1, 2003.

\bibitem{Kramer2003c}
T.~Kramer and C.~Bracher.
\newblock {\em submitted for publication}, 2003.
\newblock Online: http://arxiv.org/abs/cond-mat/0309424

\bibitem{Abramowitz1965a}
M.~Abramowitz and I.A. Stegun.
\newblock {\em Handbook of Mathematical Functions}.
\newblock Dover, New York, 1965.

\bibitem{Moellenstedt1959a}
G.~M{\"o}llenstedt and C.~J{\"o}nsson.
\newblock {\em Z. Phys.}, 155:472, 1959.

\bibitem{Demkov1982a}
Yu.N. Demkov {\em et~al}.
\newblock {\em JETP Lett.}, 34:403, 1982.

\bibitem{Fabrikant1981a}
I.I. Fabrikant.
\newblock {\em Sov. Phys. JETP}, 52:1045, 1981.

\bibitem{Bracher1999a}
C.~Bracher.
\newblock {\em Quantum Ballistic Motion and its Applications}.
\newblock PhD thesis, Technische Universit{\"a}t M{\"u}nchen, 1999.
\newblock Unpublished.

\bibitem{Kramer2002a}
T.~Kramer {\em et~al}.
\newblock {\em J. Phys. A: Math. Gen.}, 35:8361, 2002.

\bibitem{Dalidchik1976a}
F.I. Dalidchik and V.Z. Slonim.
\newblock {\em Sov. Phys. JETP}, 43:25, 1976.

\bibitem{Galilei1638a}
G.~Galilei.
\newblock {\em Discorsi e dimostrazioni matematiche intorno a due nuove scienze
  attenenti alla mecanica {\&} i movimenti locali}.
\newblock Leiden, 1638.

\bibitem{Blondel1996a}
C.~Blondel {\em et~al}.
\newblock {\em Phys.~Rev.~Lett.}, 77:3755, 1996.

\bibitem{Blondel1999a}
C.~Blondel {\em et~al}.
\newblock {\em Eur. Phys. J. D}, 5:207, 1999.

\bibitem{Kramer2001a}
T.~Kramer {\em et~al}.
\newblock {\em Europhys. Lett.}, 56:471, 2001.

\bibitem{Bracher1998a}
C.~Bracher {\em et~al}.
\newblock {\em Am. J. Phys.}, 66:38, 1998.

\bibitem{Kramer2003a}
T.~Kramer {\em et~al}.
\newblock {\em accepted for publication}, 2003.
\newblock Online: http://arxiv.org/abs/quant-ph/0307228.

\bibitem{Kramer2003b}
T.~Kramer.
\newblock {\em Matter waves from localized sources in homogeneous force
  fields}.
\newblock PhD thesis, Technische Universit{\"a}t M{\"u}nchen, 2003.
\newblock Online:  http://tumb1.biblio.tu-muenchen.de/publ/diss/ph/2003/kramer.pdf.

\end{thebibliography}

\end{document}